\documentclass[]{pasj01}
\Received{}%{yyyy/mm/dd}
\Accepted{}%{yyyy/mm/dd}
\begin{document}

\title{
%\LETTERLABEL %%% <-- uncomment for LETTER article
%\REVIEWLABEL %%% <-- uncomment for REVIEW article
Application of a two dipole model to PSR J1640-4631, a pulsar with an anomalous braking index}

%%% begin:list of authors
% Do NOT capitalize all letters in "textsc".
\author{Hui  \textsc{Shi}}
\author{Hong-Wei  \textsc{Hu}}
\author{Wen-Cong \textsc{Chen}}
\altaffiltext{}{School of Physics and Electrical Information, Shangqiu Normal University, Shangqiu 476000, China}
\email{chenwc@pku.edu.cn}

%% `\KeyWords{}' always has to be placed before ``\maketitle''
%%  List of Key Words:  https://academic.oup.com/pasj/pages/Pasj_Keywords
\KeyWords{stars: magnetic field -- stars: neutron -- stars: rotation -- pulsars: general -- pulsars: individual: PSR J1640-4631}

\maketitle

\begin{abstract}
Recent timing observation provides an intriguing result for the braking index of the X-ray pulsar PSR J1640-4631, which has a measured braking index $n=3.15\pm0.03$. The decrease of the inclination angle between between the spin axis and the magnetic axis can be responsible for such a high braking index. However, the physical mechanisms causing the change of the magnetic inclination angle have not been fully understood. In this Letter, we apply a two-dipole model given by Hamil et al (2016) to explain the decrease of the magnetic inclination angle of PSR J1640-4631. The rotation effect of a charged sphere and the magnetization of ferromagnetically ordered material produce magnetic moments $M_{1}$ and $M_{2}$, respectively. There exist a minimum of the potential energy for the magnetic moment $M_{2}$ in the magnetic field of $M_{1}$, hence the $M_{2}$ will freely rotate around the minimum energy position (i. e equilibrium position), similar to a simple pendulum. Our calculation indicate the magnetic moment $M_{2}$ would evolve towards alignment with the spin axis for PSR J1640-4631, and cause the magnetic inclination angle to decrease. The single peak in the pulse profile favors a relatively low change rate of the magnetic inclination angle.
\end{abstract}

\section{Introduction}

As the fossils of the stellar evolution, radio pulsars were thought to be rapidly rotating, strongly magnetized neutron stars,
which were the evolutionary products of the massive stars via a supernova event (Pacini 1967). The spin periods were observed to be decreasing probably due to magnetic dipole radiation, in which the radiated energy originated from the rotational energy of pulsars (Gold 1968). Assuming the spin-down of pulsars obey a power law (Lyne et al 1993)
\begin{equation}
\dot{\Omega}=-K\Omega^{n},
\end{equation}
a braking index can be defined as
\begin{equation}
n=-\frac{\Omega\ddot{\Omega}}{\dot{\Omega}^{2}},
\end{equation}
where $\Omega$, $\dot{\Omega}$, and $\ddot{\Omega}$ are the angular velocity of pulsars, the derivative and the second derivative of $\Omega$, $K$ is a proportionality constant depending on the magnetic moment of pulsars.

The braking index provides some important information concerning the evolutionary history of pulsars, hence its measurement
is very significant in understanding the spin-down process of pulsars. So far, there are nine young radio pulsars with relatively reliable braking indices (de Araujo et al. 2016b), in which eight sources have a braking index lower than three (Lyne et al. 2015), and another source PSR J1640-4631 has a braking index of $n=3.15\pm0.03$ (Archibald et al. 2016). If the rotational energy of pulsars is converted into radiation energy through pure magnetic dipole radiation and the magnetic moment of pulsars is a constant, equation (2) would yield a baking index $n=3$. Therefore, timing observations for these young radio pulsars indicate that, either other braking torques influence the spin evolution of pulsars, or the magnetic moment of pulsars changes in time.

Since the gravitational radiation of pulsar with an ellipticity could produce a braking index of $n=5$, the braking torques due to
the magnetic dipole emission combining with the gravitational radiation can result in a braking index of between three to five, which may interpret the braking index of PSR J1640-4631 (de Araujo et al. 2016a; Chen 2016). Assuming a standard magnetic diploe radiation, the spin evolution of pulsars should satisfy the following equation
\begin{equation}
I\dot{\Omega}=-\frac{2M^{2}{\rm sin}^{2}\alpha \Omega^{3}}{3c^{3}},
\end{equation}
where $I$, and $M$ are the moment of inertia, and magnetic moment of pulsars; $\alpha$ is the inclination angle between the magnetic axis and the spin axis of pulsars, $c$ is the light speed in vacuo. Using equations (2) and (3), we can derived an expression for the braking index as follows
\begin{equation}
n=3+\frac{2\Omega}{\dot{\Omega}}\left(\frac{\dot{\alpha}}{{\rm tan}\alpha}+\frac{\dot{M}}{M}\right).
\end{equation}

Because of the spin-down of pulsars, $\dot{\Omega}$ is negative. According to equation (4), a negative $\dot{\alpha}$ or $\dot{M}$ would yield a braking index greater than three. In general, the magnetic moment $M=BR^{3}$, hence the magnetic field decay can result in a braking index higher than three (Blandford \& Romani 1988; Gourgouliatos \&
Cumming 2015).  Assuming a long-term exponential decay of the magnetic field, the high baking index of PSR J1640-4631 can be explained by a combination between magnetic dipole radiation and magnetic field decay (Gao et al. 2017). Adopting a wind braking model, \cite{tong17} found the alignment of the inclination angle can affect the spin-down process of pulsars, giving rise to a braking index greater than three in the early stage and then always smaller than three. Based on the inclination angle change model, \cite{eksi16} have constrained $\alpha$ between the magnetic axis and the spin axis of PSR J1640-4631 to be $18.5\pm3$ degrees, and the change rate is $\dot{\alpha}=- (0.23\pm0.05)^{\circ}\rm ~century^{-1}$.

Some observations indicate that, the inclination angle of the Crab pulsar is increasing slowly (see also Lyne et al. 2013; Ge et al. 2016). Although the change of the inclination angle could yield a braking index that is different to three, the physical mechanisms causing $\dot{\alpha}$
have not been fully understood.  Recently, Hamil et al. (2016) proposed a novel two-dipole model (hereafter HSS model) in which the magnetic structure of pulsars includes two interacting dipole fields. In this Letter, we apply the HSS model to explain the high braking index of PSR J1640-4631.

\section{HSS model}
The magnetic field of pulsars was thought to originate from the following mechanisms: the dynamo effect (Thompson \& Duncan, 1993), and the magnetization of a ferromagnetic core in the liquid interior of neutron stars (Silverstein 1969; Pearson \& Saunier 1970). It is difficult to produce a continuous dynamo action in the pulsar because there is no internal source of energy to generate convection and activate dynamo. In this work, we consider a dipole field yielding from a rotation of a charged sphere (in principle, this dipole field is very weak, see also section 3). According to the electrodynamics, a dipole
field produced by the rotation should be coaxial and centered within the pulsar. However, the ordered material that fully located in the core of pulsars would produce a reduced and local dipole field, which is neither coaxial nor concentric with the spin of pulsars (Hamil et al. 2016).

In the HSS model, the pulsar has two dipoles with magnetic moments $M_{1}$ and $M_{2}$, and the distance between two dipole centers is a constant $r$ (see also Figure 4 of Hamil et al. 2016). The second magnetic moment $M_{2}$ is thought to be off-centered dipole (P\'{e}tri 2016), and the inclination angle between the $M_{1}$ (i. e. the spin axis) and the dipole-dipole axis is $\theta_{1}$. Because of the constant separation, the linear force between two dipoles could be ignored. The young neutron star in Cassiopeia A supernova remnant was reported to be experiencing a rapid cooling (Heinke \& Ho 2010). The rapid decline in the surface temperature probably originated from the neutron superfluidity in the core of the neutron star (Shternin et al. 2011; Page et al. 2011). Because the interior of neutron star is in superfluidity phase, the friction received by the rotational region supporting $M_{2}$
is neglected. Due to the interaction between two dipoles, the change rate of the $\theta_{2}$ between the $M_{2}$ and the dipole-dipole axis is (Hamil et al. 2016)
\begin{equation}
\dot{\theta}_{2}=\sqrt{\frac{2M_{1}M_{2}}{I_{2}r^{3}}F(\Theta)},
\end{equation}
where $F(\Theta)=({\rm sin}\theta_{1i}{\rm sin}\theta_{2i}-2{\rm cos}\theta_{1i}{\rm cos}\theta_{2i})-({\rm sin}\theta_{1f}{\rm sin}\theta_{2f}-2{\rm cos}\theta_{1f}{\rm cos}\theta_{2f})$, $I_{2}$ is the moment of inertia of the rotating region, and i and f represent the initial and
final values, respectively.

During the motion of the $M_{2}$, its potential energy would convert into the kinetic energy. Taking the constant $C=0$ in the equation (10) of Hamil et al. (2016), we can get the direction of the second dipole $M_{2}$ when its potential energy is minimum, i. e.
\begin{equation}
\theta_2^{\rm min}=-\arctan(\tan\theta_1/2).
\end{equation}
It is clear that $\theta_2^{\rm min}$ should be negative, i. e. the equilibrium position should locate the clockwise direction of the dipole-dipole axis. For an arbitrary initial $\theta_2$, the physical law indicates that the $M_{2}$ will freely rotate around the minimum energy position (i. e equilibrium position), similar to a simple pendulum.

By the geometric relation, the magnetic inclination angle $\alpha=\theta_1-\theta_2$ (noted $\theta_2<0$). After differentiating, the change rate $\dot{\alpha}$ of the magnetic inclination angle is equal to $-\dot{\theta}_{2}$. If $M_{2}$ is rotating in the vicinity of the equilibrium position with the counter clockwise direction, $\dot{\theta}_{2}>0$, hence $\dot{\alpha}<0$. When
$\dot{M}=0$, we will obtain a braking index greater than three from equation (4). On the contrary, $\dot{\theta}<0$ when $M_{2}$ is rotating in the vicinity of the equilibrium position with the clockwise direction, hence $\dot{\alpha}>0$, naturally it will result in a braking index smaller than three.

\section{Application to PSR J1640-4631}
Assuming that the high braking index of PSR J1640-4631 originated
from the change of the magnetic inclination angle, taking $\dot{M}=0$, equation (4) yields
\begin{equation}
\dot{\alpha}=(n-3){\rm tan} \alpha\frac{\dot{\Omega}}{2\Omega}.
\end{equation}
Inserting the observed parameters of PSR J1640-4631 $\Omega=30.41~\rm s^{-1}$, and
$\dot{\Omega}=-1.43\times10^{-10}~\rm s^{-2}$ (Archibald et al. 2016), in Figure 1 we plot the
relation between the braking index and the magnetic inclination angle $\alpha$ for $\dot{\alpha}=$
-0.23 (Ek\c{s}i et al. 2016), -0.56 (the absolute value is same to Crab), -1.0 (a relatively high change rate)$^{\circ}\rm ~century^{-1}$. From this relation, we can get the magntic inclination angle $\alpha\approx 75, 83, 85~^{\circ}$ for three cases, respectively. If the change rate of the magnetic inclination angle of PSR J1640-4631 is similar to the Crab, it probably has a large inclination angle.

In Table 1, we present some angle parameters of PSR J1640-4631 in the HSS model for two particular angles $\theta_{1}=45, 60^{\circ}$. When $\theta_{1}=45^{\circ}$, from equation (6) $\theta_{2}^{\rm min}=-26^{\circ}$. According to $\alpha=\theta_{1}-\theta_{2}$, we can get $\theta_{2}=-30, -38, -40^{\circ}$ for models A, B, and C, respectively. Comparing $\theta_{2}$ with $\theta_{2}^{\rm min}$, the $M_{2}$ should rotate along the counter clockwise, and the magnetic inclination angle decreases (i.e. $\dot{\alpha}<0$). However, $\theta_{2}^{\rm min}=-41^{\circ}$ when $\theta_{1}=60^{\circ}$. Because $\theta_{2}=-15, -23, -25^{\circ}$ for models D, E, and F, respectively, the $M_{2}$ should rotate along the clockwise, and the magnetic inclination angle increases (i.e. $\dot{\alpha}>0$).

To obtain a negative $\dot{\alpha}$, $\theta_{2}$ should be less than $\theta_{2}^{\rm min}$. Therefore we have
\begin{equation}
\alpha>\theta_{1}+\arctan(\tan\theta_1/2).
\end{equation}
In Figure 2, we plot the relation between the critical $\theta_{1}$ and $\alpha$, and the plane is divided into both negative $\dot{\alpha}$ and positive $\dot{\alpha}$ areas. To obtain a negative  $\dot{\alpha}$, the initial angle $\theta_{1}$ should be less than 47, 51, $52^{\circ}$ for models A, B, and C, respectively.

If the current change rate of the magnetic inclination angle of PSR J1640-4631 is same to that of the Crab pulsar, $\dot{\alpha}=- 0.56^{\circ}\rm ~century^{-1}$ (Lyne et al. 2013). Since its characteristic age is $\tau=-\Omega/2\dot{\Omega}\approx 3400~\rm yr$, the total change of $\theta_2$ is not large. For simplicity. the angular acceleration $\ddot{\theta_2}$ is assumed to approximately be a constant, hence the mean angular velocity is $\bar{\dot{\theta_2}}=- 0.28^{\circ}\rm ~century^{-1}$. Considering the characteristic age, $\theta_2$ should changed by approximately $9.5^{\circ}$ from an initial angle $\theta_{\rm 2i}$ to the current one $\theta_{\rm 2f}$ if the rotation direction has not changed.

\begin{table}
\begin{center}
\caption{Main angle parameters of PSR J1640-4631 in the two dipole model. We list the models, the change rate of the magnetic inclination angle, the magnetic inclination angle, the angle between the dipole-dipole axis and the rotation axis, the angle between the dipole-dipole axis and the $M_{2}$, and the $\theta_{2}$ when the potential energy of the $M_{2}$ is minimum. \label{tbl-2}}
\begin{tabular}{@{}llllll@{}}
\hline\hline\noalign{\smallskip}
Models    &$\dot{\alpha}$   & $\alpha$      & $\theta_{1}$ &  $\theta_{2}$ &$\theta_{2}^{\rm min}$  \\
           & ($^{\circ}\,\rm century^{-1}$)&($^{\circ}$) & ($^{\circ}$)   &  ($^{\circ}$) &  ($^{\circ}$)      \\
\hline\noalign{\smallskip}
A  & -0.23   &75 & 45   & -30 & -26\\
B  & -0.56   &83& 45   & -38 & -26\\
C  & -1.0    &85 & 45   & -40 & -26 \\
\hline\noalign{\smallskip}
D  & -0.23   &75 & 60   & -15 & -41\\
E  & -0.56   & 83& 60   & -23 & -41\\
F  & -1.0    &85 & 60   & -25 & -41 \\

\hline\noalign{\smallskip}
\end{tabular}
\end{center}
\end{table}

We take $\theta_{1,\rm i}=\theta_{1,\rm f}=45^{\circ}$, and $\theta_{2,f}=-38^{\circ}$ according to the Table 1, then $\theta_{2,i}=-47.5^{\circ}$. Based on these $\theta$ values, we can derive $F(\Theta)\approx 0.073$.
In the HSS model, the radio radiation of the pulsar come from the second magnetic moment $M_{2}$. If the PSR J1640-4631 has a standard magnetic field just like the normal pulsars, $B=10^{12}~\rm G$, and $R=10^{6}~\rm cm$, then
$M_{2}=10^{30}~\rm G\, cm^{2}$. Normally, the moment of inertia for a neutron star is $I\sim10^{45}~\rm g\,cm^{2}$ (is also the moment of inertia $I_{1}$ of $M_{1}$). Taking $I_{2}\sim10^{43}~\rm g\,cm^{2}$ (1\% of the total $I$), and $r=0.5R$, the magnetic moment $M_{1}\approx 2\times10^{9}~\rm G\, cm^{2}$. Similar to Hamil et al. (2016), we also obtain a relatively small magnetic moment for the $M_{1}$, which is consistent with its rotation origin of a charged sphere. Though $M_{1}<<M_{2}$, the $M_{2}$ should move towards its minimum potential energy in the dipole field of $M_{1}$ because $I_{1}>>I_{2}$.

\begin{figure}
\includegraphics[width=1.1\columnwidth]{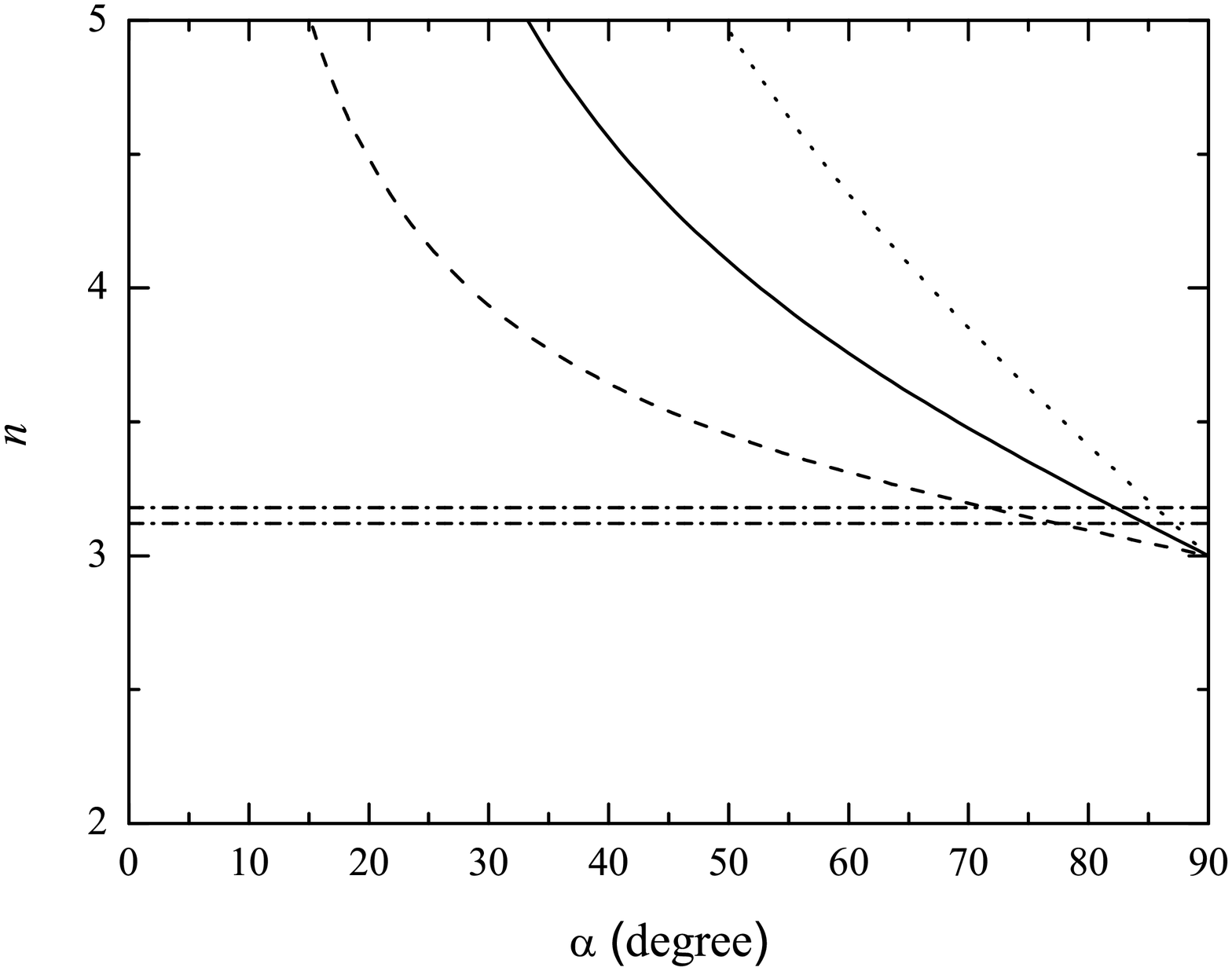}
\caption{Relation between the braking index and the magnetic inclination angle $\alpha$ for PSR J1640-4631. The solid, dashed, and dotted curves denote the cases when $\dot{\alpha}=-0.23, -0.56$, and -1.0$^{\circ}\rm ~century^{-1}$, respectively. Two horizontal dashed-dotted lines represent the possible range ($3.12-3.18$) of the measured braking index for the PSR J1640-
4631.}
\label{fig:all_model}
\end{figure}

\begin{figure}
\includegraphics[width=1.1\columnwidth]{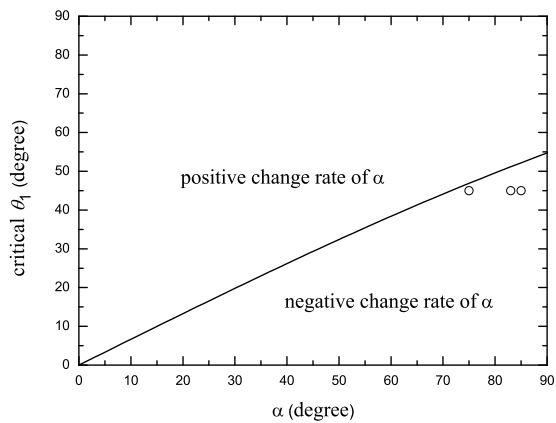}
\caption{Relation between the critical $\theta_{1}$ and $\alpha$. Three open circles represent the positions of Models A, B, and C.}
\label{fig:all_model}
\end{figure}

\section{Discussion and Summary}
Employing the HSS model, in this Letter we attempt to interpret the high braking index of PSR J1640-4631. In the HSS model, the rotation effect of a charged sphere and the magnetization of ferromagnetically ordered material produce magnetic moments $M_{1}$ and $M_{2}$, respectively. Due to the magnetic interaction, the $M_{2}$ would freely rotate around the minimum potential energy of the $M_{2}$ in the magnetic field of the $M_{1}$. The plus-minus sign of $\dot{\alpha}$ depend on the inclination angle $\theta_{1}$ between the $M_{1}$ and the dipole-dipole axis. We obtain a relation between the critical $\theta_{1}=45^{\circ}$ and the magnetic inclination angle $\alpha$ (see also Figure 2). For an angle between the dipole-dipole axis and the rotation axis $\theta_{1}=45^{\circ}$, the calculated $\theta_{2}$ indicate that the $M_{2}$ should rotate along the counter clockwise, and cause a negative $\dot{\alpha}$ and a braking index greater than three. On the contrary, a $\theta_{1}=60^{\circ}$ would result in a positive $\dot{\alpha}$ and a braking index less than three. In a word, the HSS model can account for a positive or negative $\dot{\alpha}$, and a braking index greater of less than three.

In principle, an orthogonal rotator would typically produce emission when each pole aim at the earth, so the single peak in the pulse profile tends to a small inclination angle (Rankin 1983, 1990; Weltevrede \& Johnston 2008; Hankins \&
Rankin 2010). Therefore, PSR J1640-4631 may have a relatively low change rate (see also model A in Table 1) of the inclination angle comparing with the Crab if its anomalous braking index arose from the change of the magnetic inclination angle.

Yi \& Zhang (2015) proposed that the long-term red timing-noise caused by the evolution of the magnetic inclination
angle of pulsars. Because the HSS model could be alternative explanation in understanding the evolution of the magnetic inclination angle, the free swing of a dipole in the magnetic field of another dipole could be a possible physical mechanism resulting in the long-term red timing-noise.

According to HSS model, the magnetic moment $M_{2}$ of PSR J1640-4631 would evolve towards alignment with the spin axis, and its potential energy would gradually decrease. From equation (11) of Hamil et al. (2016), the change rate of the magnetic inclination angle should increase. However, PSR J1640-4631 is a radio quiet pulsar (Archibald et al. 2016), and had not been detected gamma-ray pulsations from PSR J1640-4631 (Gotthelf et al. 2014), hence there is no the observed results for the value of $\alpha$. We expect further observations aiming at the gamma-ray or radio pulse profiles to constrain the magnetic inclination angle of this source. Recently, Cheng et al. (2019) proposed that future observation of the magnetic inclination angle could help us to constrain the number of precession cycles, and provide information about the internal magnetic field configuration of PSR J1640-4631.

\begin{ack}
We thank the referee for a very careful reading and
comments that have led to the improvement of the manuscript. This work was partly supported by the National Natural Science Foundation of China (under grant number 11573016, 11733009), the Program for Innovative Research Team (in Science and Technology) at the University of Henan Province.
\end{ack}

%%%
% See the manual for the detail.
%%%

\end{document}